\documentclass[conference]{IEEEtran}
\IEEEoverridecommandlockouts
\usepackage{cite}
\usepackage{amsmath,amssymb,amsfonts}
\usepackage{graphicx}
\usepackage{textcomp}
\usepackage{xcolor}
\def\BibTeX{{\rm B\kern-.05em{\sc i\kern-.025em b}\kern-.08em
    T\kern-.1667em\lower.7ex\hbox{E}\kern-.125emX}}

\usepackage{bm}  
\usepackage{multicol}
\usepackage{multirow}
\usepackage{placeins}
\usepackage{graphicx}
\usepackage{amsmath}
\usepackage{array}
\newcolumntype{H}{>{\setbox0=\hbox\bgroup}c<{\egroup}@{}}
\usepackage{graphicx}
\usepackage{caption}
\usepackage{subcaption}
\usepackage{float}
\usepackage{tabularx}
\usepackage{arydshln}
\usepackage[autostyle]{csquotes}
\usepackage{color}
\usepackage{amssymb}

\usepackage[utf8]{inputenc}
\usepackage[english]{babel}

\usepackage{amsthm}

\usepackage[hidelinks]{hyperref}
\usepackage{algorithm}
\usepackage{algpseudocode}

\usepackage{chngpage}
\usepackage{balance} % \usepackage{balance} in the beginning of the latex document, and then add \balance somewhere in the left column text of the last page.

\usepackage{tablefootnote}

\newtheorem{conclusion}{Conclusion}

\begin{document}

\title{Solving the Steiner Tree Problem in Graphs \\using Physarum-inspired Algorithms}

\author{\IEEEauthorblockN{Yahui Sun}
	\IEEEauthorblockA{\url{https://yahuisun.com}}
	}

\maketitle

\begin{abstract}
Some biological experiments show that the tubular structures of Physarum polycephalum are often analogous to those of Steiner trees. Therefore, the emerging Physarum-inspired Algorithms (PAs) have the potential of computing Steiner trees. In this paper, we propose two PAs to solve the Steiner Tree Problem in Graphs (STPG).  We apply some widely-used artificial and real-world VLSI design instances to evaluate the performance of our PAs. The experimental results show that: 1) for instances with hundreds of vertices, our first PA can find feasible solutions with an average error of 0.19\%, while the Genetic Algorithm (GA), the Discrete Particle Swarm Optimization (DPSO) algorithm  and a widely-used Steiner tree approximation algorithm: the Shortest Path Heuristic (SPH) algorithm can only find feasible solutions with an average error above 4.96\%; and 2) for larger instances with up to tens of thousands of vertices, where our first PA, GA and DPSO are too slow to be used, our second PA can find feasible solutions with an average error of 3.69\%, while SPH can only find feasible solutions with an average error of 6.42\%. These experimental results indicate that PAs can compute Steiner trees, and it may be preferable to apply our PAs to solve STPG in some cases. 
\end{abstract}


\section{Introduction}
Physarum polycephalum is a slime mold that inhabits shady, cool and moist areas. It has exhibited many intelligent behaviors, such as maze solving and efficient-network building \cite{imba2000, pmer2007, snsi2004,smeo2012,r2a72014}. To exploit its ability, Physarum polycephalum has been studied intensively over the last few decades, and several Physarum-inspired Algorithms (PAs) have been proposed to solve network optimization problems, such as the shortest path problem \cite{fspo2012}, and the traveling salesman problem \cite{auos2014}.

The Steiner Tree Problem in Graphs (STPG) is a well-known NP-hard network optimization problem that has been applied to VLSI design \cite{RST2015,pspf2002,stst2018}, communication network design \cite{tfha2019,mhnp2019}, system biology \cite{apps2016,tnst2017}, and knowledge management \cite{faos2012,sstf2012}. Let $G(V,E,c)$ be a connected undirected graph, where $V$ is the set of vertices, $E$ is the set of edges, and $c$ is a function which maps each edge in $E$ to a positive number called the edge length. Let $T$ be a subset of $V$ called terminals. STPG aims to find a connected subgraph $G'\subseteq G$ that contains all the terminals for which the sum of edge lengths in $G'$ is minimum.  The subgraph $G'$, which is an optimal solution to STPG, is called a Steiner Minimum Tree (SMT) in $G$ for $T$. Due to the NP-hardness, the time required to find an SMT may increase exponentially as the graph size increases. However, many real-world instances for STPG involve large graphs with thousands or even tens of thousands of vertices \cite{saul2001,assk2012,pdim2010}. Thus, it is necessary to develop algorithms that have good performances in large graphs.

On the other hand, some biological experiments have shown that the tubular structures of Physarum polycephalum are often analogous to those of Steiner trees (e.g. \cite{omsf2004}), which means that PAs may have the potential of computing Steiner trees \cite{faip2008,amib2010, poab2015,faib2016,apps2016}. There have already been some attempts to propose PAs to solve STPG \cite{faip2008,amib2010,poab2015}. Nevertheless, most of them can only find approximations to SMTs in small graphs \cite{faip2008,amib2010}, while others that can be used in large graphs can only be used in graphs with equal edge length \cite{poab2015}, which restricts their usage in reality. Moreover, none of the existing PAs has shown a better performance than any of the most widely-used algorithms for STPG. To our knowledge, no PA has even been tested for well-known benchmark instances so far. In this paper, we will address this issue by proposing two new PAs for STPG, and comparing them with several widely-used algorithms for some well-known benchmark instances. Below are our main contributions. 

\begin{itemize}

	\item we propose two new PAs: the Multiple Sources Single Sink Physarum Optimization (MS$^3$-PO) algorithm and the Hierarchical Multiple Sources Single Sink Physarum Optimization (HMS$^3$-PO) algorithm to solve STPG.
	
	\item we indicate the competitiveness of our new PAs over the Genetic Algorithm (GA), the Discrete Particle Swarm Optimization (DPSO) algorithm  and a Steiner tree approximation algorithm: the Shortest Path Heuristic (SPH) algorithm \cite{aasf1980} for some widely-used artificial and real-world VLSI design instances\protect\footnotemark.
\end{itemize}

\addtocounter{footnote}{0}\footnotetext{\textbf{The codes and  datasets are available}  at https://github.com/YahuiSun/PAs-for-Steiner-Tree-Problem-in-Graphs}

\begin{figure}	[!t]
	\vspace{0.1cm}
	\centering
	\begin{subfigure}[t]{0.475\textwidth}
		\centering
		\includegraphics[width=\textwidth,height=0.12\textheight]{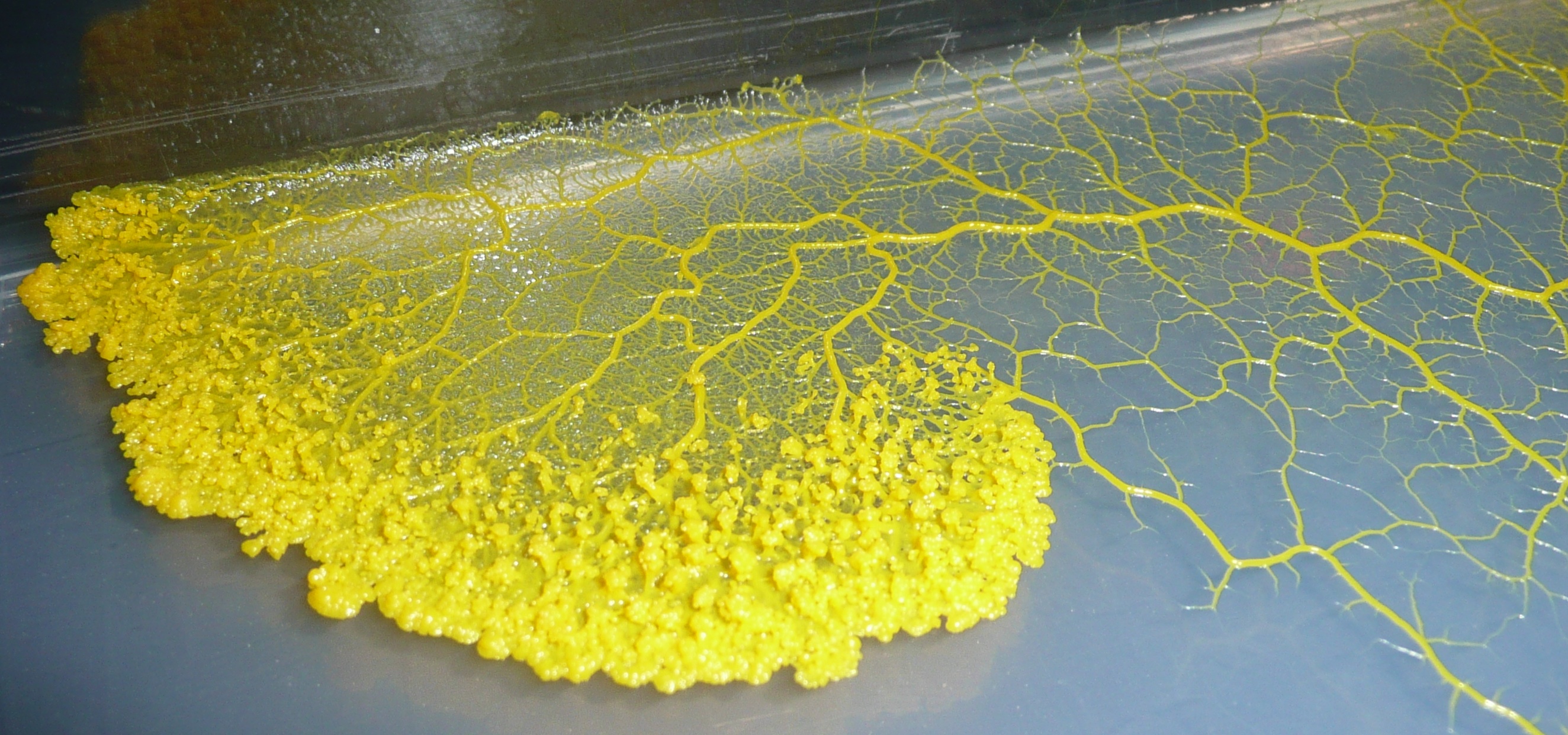}
		\subcaption{The tubular structure of Physarum polycephalum}
		\label{ppp1}	
	\end{subfigure}
	\hfill
	\begin{subfigure}[t]{0.475\textwidth}
		\centering
		\includegraphics[width=\textwidth]{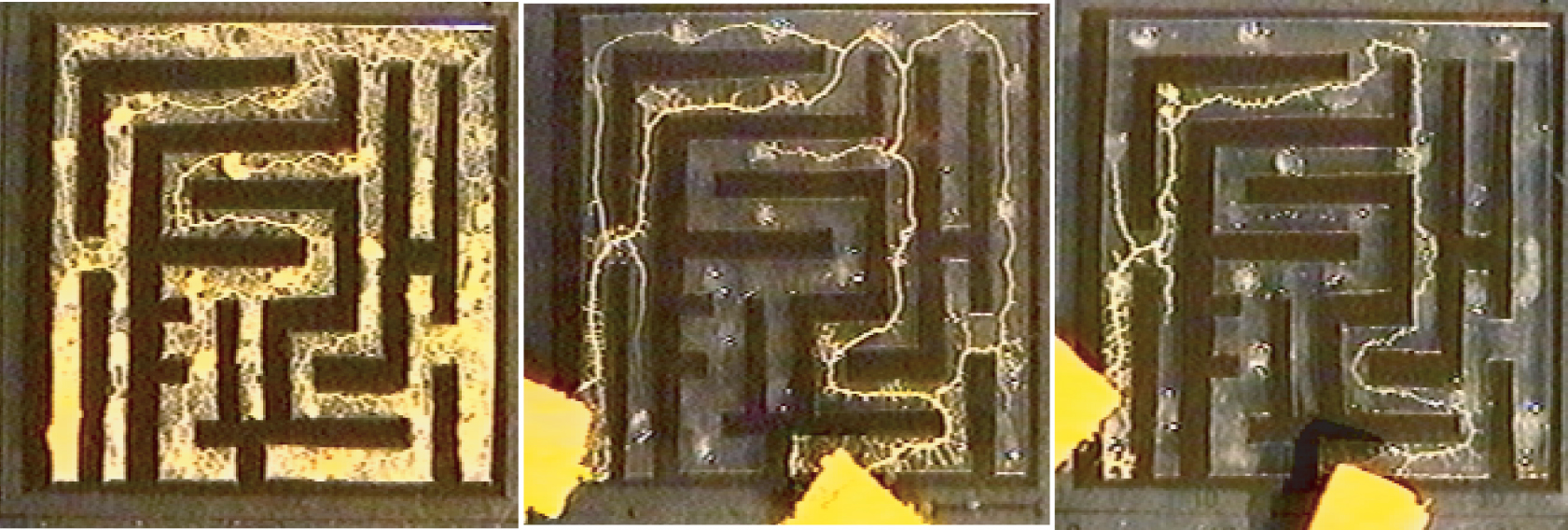}
		\subcaption{Physarum polycephalum finds the shortest path}
		\label{ppp2}
	\end{subfigure}
	\caption{Photographs of Physarum polycephalum (provided by Professor Toshiyuki Nakagaki from Hokkaido University). (a) shows the tubular structure of Physarum polycephalum. (b) shows an experiment in which Physarum polycephalum finds the shortest path between two agar blocks in a maze \cite{imba2000}.}
	\label{ppp photo}
\end{figure}

\section{Related work: Physarum Solver for the shortest path problem}

Besides the attempts to solve STPG, some PAs have already solved the shortest path problem \cite{pccs2012}. A well-known one of such PAs is the Physarum Solver \cite{psab2006}, which is the basis of many PAs \cite{faip2008,poab2015,faib2016}. We introduce it in this section. 

In a previous experiment, Physarum polycephalum has demonstrated the ability of finding the shortest path between two agar blocks in a maze \cite{imba2000}. This path-finding ability is attributed to an underlying physiological mechanism: Physarum's tube thickens as the protoplasmic flux through it increases \cite{ammf2007,sbot2001}. Physarum Solver is inspired by the mechanism above. In Physarum Solver, the graph is considered as a network with protoplasmic flux in every edge. The two terminals in the shortest path problem represent two agar blocks containing nutrient, which are food for Physarum polycephalum. One terminal is called the source node, and the other terminal is called the sink node. The protoplasmic flux flows into the graph from the source node and out of the graph from the sink node. There is pressure at each vertex, and the quantity of flux in each edge is proportional to the pressure difference between two ends of this edge. Specifically, the flux $Q_{ij}$ in edge $(i,j)$ is given by the Hagen-Poiseuille equation below,
\begin{eqnarray} \label{calQ}
Q_{ij} = \frac{D_{ij}}{c_{ij}} (p_i - p_j)
\end{eqnarray}
\begin{eqnarray} \label{D_r}
D_{ij} = \frac{\pi r^4_{ij}}{8 \xi}
\end{eqnarray}
\noindent where $D_{ij}$ is the edge conductivity, $c_{ij}$ is the edge length, $p_i$ and $p_j$ are pressures at vertex $i$ and $j$, $r_{ij}$ is the edge radius, and $\xi$ is the viscosity coefficient. Equation (\ref{D_r}) shows that Physarum's tubular thickness ($r_{ij}$) increases with the tube's conductivity. Thus, the change of Physarum's tubular thickness can be described by the conductivity update equation as follows,
\begin{eqnarray} \label{updateD}
\frac{d}{dt} D_{ij} = \alpha|Q_{ij}| - \mu D_{ij}
\end{eqnarray}
where $\alpha$ and $\mu$ are two positive constants. The conductivity update equation implies that conductivities tend to increase in edges with large flux. Therefore, the conductivity update equation reflects the physiological mechanism that Physarum's tube thickens as the protoplasmic flux through it increases. To calculate the flux and update edge conductivities, we need to first calculate the pressures. By considering the conservation law of flux at each vertex, the pressures can be calculated using the network Poisson equation below,
\begin{eqnarray} \label{old pressure}
\sum_{i \in V(j)} \frac{D_{ij}}{c_{ij}} (p_i - p_j) =\left\{
\begin{array}{c l}      
-I_0, & j=source\\
+I_0, & j=sink\\
0, & otherwise
\end{array}\right.
\end{eqnarray}
\noindent where $V(j)$ is the set of vertices linked to vertex $j$, and $I_0$ is the quantity of flux flowing into the source node and out of the sink node. Let the pressure at the sink node be 0, and give each edge conductivity an initial value; then the other pressures can be calculated using Equation (\ref{old pressure}). After that, the flux in each edge can be calculated using Equation (\ref{calQ}), and the conductivity of each edge can be updated using Equation (\ref{updateD}). It is easy to see that the pressures (excluding the pressure at the sink node) and the fluxes will also change after the update of edge conductivities. 

Use $\epsilon$ to signify the threshold value of edge conductivity. Edges with conductivities smaller than $\epsilon$ will be cut from the network. Ultimately, if there is a unique shortest path between the source node and the sink node, then this path can be found by iteratively updating edge conductivities and cutting edges \cite{pccs2012}. However, even though Physarum Solver can solve the shortest path problem, it cannot solve STPG directly because there are only two terminals in its model, which are the source node and the sink node, while there are multiple terminals in STPG. Thus, new PAs are required to solve STPG.

\section{The proposed Multiple Sources Single Sink Physarum Optimization algorithm}

The major difference between STPG and the shortest path problem is that there may be more than two terminals in STPG. Therefore, one of the challenges of using PAs to solve STPG is to select  multiple terminals to be source or sink nodes. There are two ways to do this. One way is to select two terminals to be a single source node and a single sink node \cite{plan2013, toir2011,caoc2008}. The other way is to select one terminal to be a single sink node and then select all the other terminals to be multiple source nodes \cite{poab2015,otso2015}.  The Multiple Sources Single Sink Physarum Optimization (MS$^3$-PO) algorithm is proposed in this section to solve STPG. In MS$^3$-PO, one terminal is selected from all the terminals at a time to be the sink node, and all the other terminals will be selected to be source nodes. 

We propose a probabilistic method to select one terminal at a time to be the sink node. Let $l(i)$ be the total length of edges incident to terminal $i$. Name the terminals in such a way that $l(1) \leq l(2) \leq \cdots \leq l(|T|)$, where $|T|$ is the number of terminals. Then, we set the probability of selecting terminal $i$ to be the sink node by  
\begin{eqnarray}\label{probablity}
P(i)= \frac{l(|T|-i+1)}{\sum_{j=1}^{|T|} l(j)}
\end{eqnarray}

It can be seen from Equation (\ref{probablity}) that the larger the total edge length incident to a terminal, the smaller the probability that it is selected to be the sink node. The logic behind Equation (\ref{probablity}) is that a terminal associated with a larger total edge length is more likely to join unwanted edges. Moreover, because the flux in each source node is $I_0$ and the flux in the sink node is $(|T|-1)I_0$, it is easier to cut edges incident to a source node than to cut edges incident to the sink node ($(|T|-1)I_0 \geq I_0$). Therefore, it is preferable to select terminals associated with large total edge lengths to be source nodes.

After selecting a single sink node and $|T|-1$ source nodes, the network Poisson equation can be adapted from Equation (\ref{old pressure}) to
\begin{eqnarray} \label{getQ}
\sum_{i \in V(j)} \frac{D_{ij}}{c_{ij}} (p_i - p_j) =\left\{
\begin{array}{c l}      
-I_0, & j=source\\
+(|T|-1)I_0, & j=sink\\
0, & otherwise
\end{array}\right.
\end{eqnarray}

Once the pressure at the sink node is set to 0, all the other pressures can be calculated by solving the adapted network Poisson equation. Then, the flux through each edge can be calculated using Equation (\ref{calQ}). Ultimately, SMTs or close approximations to SMTs can be found by iteratively updating edge conductivities and cutting edges in the same way as Physarum Solver. 

Due to the randomness in selecting the sink node, MS$^3$-PO has the ability to find different feasible solutions for a single instance. Therefore, it is recommended to find as many solutions as possible and then select the best one as the final solution. However, it is impossible to obtain more than one solution without retrieving the edges that have already been cut from the initial graph. Thus, the graph and the edge conductivities need to be initialized after a certain number of conductivity update times. In MS$^3$-PO, a parameter $K$ is used as the upper limit of $k$, which is the number of conductivity update times. After $K$ times of conductivity update, a feasible solution can be found. Then, the graph and edge conductivities will be initialized to find another feasible solution. A parameter $M$ is used as the upper limit of $m$, which is the number of initialization times. The iteration of conductivity update is called the inner iteration process, and the iteration of graph initialization is called the outer iteration process.

In Physarum Solver, the conductivity update equation (Equation (\ref{updateD})) can also be written as
\begin{eqnarray}\label{oldC}
D_{ij}(k+1)=D_{ij}(k)+\alpha|Q_{ij}(k)|-\mu D_{ij}(k)
\end{eqnarray}
where $k$ is the number of conductivity update times, and $k \geq 1$. While in MS$^3$-PO, we propose an evolutionary computation technique to update edge conductivities, and two new equations are used in this technique, 
\begin{eqnarray}\label{newC1}
\begin{split}
D_{ij}(k+1)=(1+\delta)*[D_{ij}(k)+\alpha|Q_{ij}(k)|-\mu D_{ij}(k)]
\end{split}
\end{eqnarray}
\begin{eqnarray}\label{newC2}
\begin{split}
D_{ij}(k+1)=(1-\delta)*[D_{ij}(k)+\alpha|Q_{ij}(k)|-\mu D_{ij}(k)]
\end{split}
\end{eqnarray}
\noindent where $\delta$ is a positive number. MS$^3$-PO can find different feasible solutions through different outer iterations, and the solution network with the smallest total edge length is saved as the best network found so far (initially, the saved network is graph $G$). If edge ($i,j$) belongs to the saved network, its conductivity will be updated using Equation (\ref{newC1}), or its conductivity will be updated using Equation (\ref{newC2}). It can be seen that Equation (\ref{newC1}) gives edge ($i,j$) a bigger conductivity and makes it survive longer, while Equation (\ref{newC2}) gives edge ($i,j$) a smaller conductivity and makes it survive less long. The computational trials show that this technique can accelerate the optimization process of MS$^3$-PO. 

On the other hand, there is no guarantee that all the not-cut-yet edges are connected. Therefore, it is necessary to cut the edges which are not connected with the terminals after each inner iteration.  Moreover, there is also no guarantee that the solution network of MS$^3$-PO is a tree. Thus, we find the Minimum Spanning Tree (MST) of the saved network after the outer iteration process to ensure that the solution network of MS$^3$-PO is a tree. This MST is considered as the final solution of MS$^3$-PO.

\begin{algorithm}[!h] 
	\caption{The proposed Multiple Sources Single Sink Physarum Optimization algorithm (MS$^3$-PO)}
	\textbf{Input:} Graph $G(V,E,c)$, terminal set $T$, and parameters $M$, $K$, $\delta$, $\alpha$, $\mu$, $\epsilon$\\
	\textbf{Output:} Steiner tree $G' \subset G$ 
	\begin{algorithmic}[1]
		\State Initialize $G'=G$
		\For {$m$ = $1$ to $M$}
		\State Initialize $G$ and the edge conductivities
		\For {$k$ = $1$ to $K$}
		\State Select source and sink nodes using Equation (\ref{probablity})
		\State Calculate pressures using Equation (\ref{getQ})
		\State Calculate fluxes using Equation (\ref{calQ})
		\State Update conductivities using Equation (\ref{newC1}) \& (\ref{newC2})
		\State Cut edges using the threshold value $\epsilon$
		\State Cut edges that are disconnected with the terminals
		\If {$c(G')>c(G)$}
		\State $G'=G$
		\EndIf
		\EndFor
		\EndFor
		\State $G'=MST(G')$
	\end{algorithmic}
	\label{Algorithm: MS3-PO} % it should be below \caption
\end{algorithm}

\begin{algorithm}[!b] 
	\caption{The proposed Hierarchical Multiple Sources Single Sink Physarum Optimization algorithm (HMS$^3$-PO)}
	\textbf{Input:} Graph $G(V,E,c)$, terminal set $T$, and parameters $M$, $K$, $\delta$, $\eta$, $\mu$, $\epsilon$, $d$, $n$\\
	\textbf{Output:} Steiner tree $G' \subset G$
	\begin{algorithmic}[1] 
		\State Let each terminal be a subset
		\While {there are vertices not included by the subsets}
		\State Calculate the shortest distance between each subset and each not-included-yet vertex
		\State Find the smallest shortest distance and the corresponding subset and vertex
		\If {the corresponding subset contains no more than $d|V|/n-1$ vertices}
		\State Add the corresponding vertex to the subset
		\Else 
		\State Break
		\EndIf
		\EndWhile
		\While {there are more than $n$ subsets}
		\State Find the subset with the least number of vertices and its closest neighbor (subset)
		\If {the total number of vertices in these two subsets and the shortest path between them is no more than $d|V|/n$}
		\State Merge these two subsets and the vertices on the shortest path between them as a new subset
		\Else
		\State Break
		\EndIf
		\EndWhile
		\State Construct subgraphs using the subsets obtained above
		\State Use MS$^3$-PO to solve STPG in each subgraph
		\State Mark all the solution networks in $G$ as connected networks
		\State Find the MST of each connected network
		\State Merge the vertices in each connected network as a new terminal
		\State Construct a new graph using the new terminals and the vertices that are not in the MSTs above
		\State Randomly choose a new terminal as the start vertex of the SPH tree
		\While {not all the new terminals have been added}
		\State Find the new terminal which is closest to the SPH tree
		\State Add the shortest path between the new terminal and the SPH tree to the SPH tree
		\EndWhile
		\State Combine the MSTs and the SPH tree together as $G'$
	\end{algorithmic}
	\label{HMS$^3$-PO Algorithm} % it should be below \caption
\end{algorithm}

\section{The proposed Hierarchical Multiple Sources Single Sink Physarum Optimization algorithm}

The proposed MS$^3$-PO above can find SMTs or close approximations to SMTs in graphs with hundreds of vertices. However, many real-world instances involve larger graphs \cite{saul2001} where it is too slow to apply MS$^3$-PO. Therefore, in this section, the Hierarchical Multiple Sources Single Sink Physarum Optimization  (HMS$^3$-PO) algorithm is proposed to address this issue. 

There are three stages in HMS$^3$-PO. In Stage 1, the large graph is partitioned into several smaller subgraphs, and each subgraph contains at least one terminal. The graph-partitioning algorithm used in HMS$^3$-PO is adapted from an algorithm proposed by Leitner et al. \cite{aphf2014}. In our adapted algorithm, $|T|$ subsets of connected vertices are first identified from the graph, where $|T|$ is the number of terminals. Each of these subsets contains a terminal. Then, these subsets are merged together as some larger subsets. Each of these merged subsets contains no more than $d|V|/n$ vertices, where $d$ is the imbalance parameter and $d \geq 1$, $|V|$ is the number of vertices, $n$ is the target number of partitioned subgraphs, and $n \leq |T|$. At last, subgraphs are constructed based on these subsets of vertices. To maintain the connectivity of each subgraph, the vertices on the shortest path between two subsets are also merged when the two subsets are merged together. Therefore, it is possible that a vertex belongs to multiple subgraphs at the same time. Moreover, since there is a limitation to the number of vertices in each subgraph, the number of partitioned subgraphs may not equal the target number of partitioned subgraphs ($n$). 

In Stage 2, each subgraph is considered as an independent instance for STPG, and MS$^3$-PO is used to find SMTs or close approximations to SMTs in these subgraphs. If a subgraph contains only one terminal, the solution in this subgraph will be the single terminal.

In Stage 3, all the solution networks obtained in Stage 2 are marked as connected networks in graph $G$. Remarkably, since a vertex may belong to different solution networks, the connected network above may not be a tree. Thus, the MST of each connected network is found. To connect these MSTs together, the vertices in each MST are merged together as a new terminal. A new graph is constructed using these new terminals and the non-terminal vertices that are not in any of the MSTs above. In the new graph, a non-terminal vertex is connected with a new terminal when this vertex is connected with at least one vertex in the corresponding MST in the initial graph, and the edge length between this vertex and the new terminal is the shortest edge length between this vertex and the vertices in the corresponding MST. 

Ultimately, a widely-used Steiner tree algorithm: the Shortest Path Heuristic (SPH) algorithm \cite{aasf1980}, is used to connect these new terminals together in the new graph. The final solution network of HMS$^3$-PO contains both the edges in the MSTs that correspond to these new terminals and the edges that are used to connect the new terminals. HMS$^3$-PO is the combination of graph-partitioning algorithm, MS$^3$-PO, and SPH. Given that the time required by MS$^3$-PO to solve STPG in the partitioned subgraphs is significantly decreased, HMS$^3$-PO solves STPG much faster than MS$^3$-PO in large graphs.

\begin{table*}[!t]
	\renewcommand{\arraystretch}{1.3}
	\caption{The comparison in the B instances}
	\label{Comparison of MS$^3$-PO, GA and DPSO in B instances}
	\centering
	\begin{tabular}{l c c c c c c c c c c}
		\hline
		\multicolumn{1}{c|}{\multirow{2}{*}{Instance}} & \multicolumn{1}{c|}{\multirow{2}{*}{$|V|$}} & \multicolumn{1}{c|}{\multirow{2}{*}{$|E|$}} & \multicolumn{1}{c|}{\multirow{2}{*}{$|T|$}} & \multicolumn{3}{c|}{Fitness Evaluation Times} & \multicolumn{4}{c}{Error/\%} \\
		\cline{5-11}
		\multicolumn{1}{c|}{}&\multicolumn{1}{c|}{}&\multicolumn{1}{c|}{}&\multicolumn{1}{c|}{}& MS$^3$-PO & GA & \multicolumn{1}{c|}{DPSO} & MS$^3$-PO & GA & DPSO & SPH\\
		\hline
		B1 & 50 & 63 & 9 & {23} & 549 & 125490 & 0 & 0 & 0 & 0\\
		B2 & 50 & 63 & 13 & {300} & 17696 & 6451 & 0 & 0 & 0 & 8.43\\
		B3 & 50 & 63 & 25 & 3978 & 156 & 2174 & 0 & 0 & 0 & 1.45\\
		B4 & 50 & 100 & 9 & {588} & 11385 & 8566 & 0 & 0 & 0 & 0\\
		B5 & 50 & 100 & 13 & 1136 & 487 & 22828 & 0 & 0 & 0 & 4.92\\
		B6 & 50 & 100 & 25 & 5250 & 128 & 10258 & 0 & 0 & 0 & 4.10\\
		B7 & 75 & 94 & 13 & {2304} & 3532 & 565720 & 0 & 0 & 0 & 0\\
		B8 & 75 & 94 & 19 & 2080 & 720 & 167061 & 0 & 0 & 0 & 0\\
		B9 & 75 & 94 & 38 & 14896 & 450 & 26912 & 0 & 0 & 0 & 1.82\\
		B10 & 75 & 150 & 13 & {7258} & 101694 & 826078 & 0 & 0 & 0 & 13.95\\
		B11 & 75 & 150 & 19 & {44591} & 159459 & 131341 & 0 & 0 & 0 & 5.68\\
		B12 & 75 & 150 & 38 & 121880 & 366 & 4879 & 0 & 0 & 0 & 0\\
		B13 & 100 & 125 & 17 & {100848} & 197863 & 1000000 & 0 & 0 & 39.39 & 6.06\\
		B14 & 100 & 125 & 25 & {28084} & 1000000 & 1000000 & 0 & 1.28 & 60.43 & 0.85\\
		B15 & 100 & 125 & 50 & 1000000 & 102418 & 1000000 & 0 & 0 & 0.63 & 1.26\\
		B16 & 100 & 200 & 17 & {36960} & 1000000 & 1000000 & 0 & 5.51 & 17.32 & 7.87\\
		B17 & 100 & 200 & 25 & 14405 & 1436 & 1000000 & 0 & 0 & 3.05 & 1.53\\
		B18 & 100 & 200 & 50 & 1000000 & 1014 & 351238 & 0.46 & 0 & 0 & 2.29\\
		\hline
	\end{tabular}
\end{table*}

\section{Experimental evaluation}

In this section, we indicate the competitiveness of our new PAs through experiments. 

\noindent{\underline{Algorithms:}} Five algorithms have been implemented: 1) MS$^3$-PO; 2) HMS$^3$-PO; 3) the Genetic Algorithm (GA) \cite{stgs1993}; 4) the Discrete Particle Swarm Optimization (DPSO) algorithm  \cite{apso2010}; and 5) a widely-used Steiner tree approximation algorithm: the Shortest Path Heuristic (SPH) algorithm \cite{aasf1980,stmf2013,aphf2014,phft1992}.

\noindent{\underline{Benchmarks:}} Two types of benchmark instances are applied: 1) the B instances from OR-library, which are used as benchmarks by many researchers \cite{saul2001,stgs1993,ofsv1986,asaf1989}; and 2) some real-world VLSI design instances from SteinLib \cite{saul2001}.

\noindent{\underline{Metrics:}} Two metrics are used for the comparison: 1) the solution error, which is defined as the difference between the sum of edge lengths in the solution found by each algorithm and the sum of edge lengths in an SMT; and 2) the fitness evaluation times, which are the number of conductivity update times in MS$^3$-PO, the number of crossover times in GA, and the number of particle position update times in DPSO. The fitness evaluation times may be a fairer metric than the CPU running time as it does not depend on machine, operating system, software environment, or coding \cite{boaa2014}. The fitness evaluation times needed by MS$^3$-PO, GA and DPSO to find an SMT have been recorded, and the upper limit of fitness evaluation times is 1 million. If an algorithm has not found an SMT in 1 million fitness evaluation times, the recorded fitness evaluation times of this algorithm will be 1 million. Remarkably, in MS$^3$-PO, the MST of the saved network is only found after 1 million fitness evaluation times (the last step of MS$^3$-PO). If an SMT has already been found by MS$^3$-PO within 1 million fitness evaluation times, which means that the saved network in MS$^3$-PO is an SMT, then there is no need to find the MST of the saved network any more. Note that, since there is no fitness evaluation in SPH, only the solution error is used as the metric in the comparison between SPH and the other algorithms. We consider that an algorithm has a better performance than the other algorithms when it needs a smaller fitness evaluation times to find an SMT, or when its solution has a smaller error than the solutions of the other algorithms.

\subsection{Applying MS$^3$-PO to the B instances}

Here, we apply MS$^3$-PO to the B instances. The comparison results are listed in Table \ref{Comparison of MS$^3$-PO, GA and DPSO in B instances}, in which $|V|$ is the number of vertices, $|E|$ is the number of edges, and $|T|$ is the number of terminals. The parameter settings of MS$^3$-PO, GA and DPSO are: in MS$^3$-PO, $I_0=1$, $\epsilon =0.001$, $\alpha =0.15$, $\mu =1$, $\delta=0.2$; in GA, the population size is 10, the mutation probability of each chromosome is 0.2; in DPSO, the population size is 10, the neighborhood radius is 2, and the mutation probability of the best position is 0.1. The parameter settings above are the best settings found in our experiments. %Notably, the terminals are always connected in the solutions of MS$^3$-PO with the parameter settings above, even though the conditions of Theorem 1 are not met ($\epsilon > \frac{4\alpha I_0}{|V|^2}$; we give $\epsilon$ a big value to make MS$^3$-PO fast).

It can be seen from Table \ref{Comparison of MS$^3$-PO, GA and DPSO in B instances} that MS$^3$-PO finds SMTs in 17 instances (B1-B17). On the contrary, GA finds SMTs in 16 instances (B1-B13, B15, B17, B18), DPSO finds SMTs in 13 instances (B1-B12, B18), while SPH can only find SMTs in 5 instances (B1, B4, B7, B8, B12). Therefore, MS$^3$-PO has an advantage over GA, DPSO and SPH by finding SMTs in more instances. Note that, the SMT in B15 is not found by MS$^3$-PO within 1 million fitness evaluation times, but is found by MS$^3$-PO as the MST of the saved network after 1 million fitness evaluation times.

\begin{table}[!b]
	\vspace{0.15cm}
	\renewcommand{\arraystretch}{1.3}
	\caption{The required fitness evaluation times of MS$^3$-PO to find SMTs with different conductivity update equations}
	\label{value of new parameter}
	\centering
	\begin{tabular}{c|c|c|c|c|c|c}
		\hline
		Eq. & B1 & B2 & B3 & B4 & B5 & B6\\
		\hline
		(\ref{newC1}) \& (\ref{newC2}) & 23 & 300 & 3978 & 588 & 1136 & 5250\\
		(\ref{oldC}) & 36 & 301 & 258778 & 594 & 1820 & 205983\\
		\hline
	\end{tabular}
\end{table}

\begin{table*}[!t]
	\renewcommand{\arraystretch}{1.3}
	\caption{The comparison in small VLSI design instances}
	\label{Comparison of MS$^3$-PO, GA and DPSO in VLSI instances}
	\centering
	\begin{tabular}{l c c c c c c c c c c c c c}
		\hline
		\multicolumn{1}{l|}{\multirow{2}{*}{Instance}} & \multicolumn{1}{c|}{\multirow{2}{*}{$|V|$}} & \multicolumn{1}{c|}{\multirow{2}{*}{$|E|$}} & \multicolumn{1}{c|}{\multirow{2}{*}{$|T|$}} & 
		\multicolumn{3}{c|}{Fitness Evaluation Times} & \multicolumn{4}{c}{Error/\%} \\
		\cline{5-11}
		\multicolumn{1}{c|}{}&\multicolumn{1}{c|}{}&\multicolumn{1}{c|}{}&\multicolumn{1}{c|}{}& MS$^3$-PO & GA & \multicolumn{1}{c|}{DPSO} & MS$^3$-PO & GA & DPSO & SPH\\
		\hline
		DIW0393 & 212 & 381 & 11  & 1634 & 1000000 & 1000000 & 0 & 37.75 & 232.12 & 0\\
		DIW0460 & 339 & 579 & 13 & 230868 & 1000000 & 1000000 & 0 & 55.07 & 432.46 & 14.20\\
		DIW0540 & 286 & 465 & 10  & 4224 & 1000000 & 1000000 & 0 & 49.73 & 326.47 & 2.14\\
		DMXA0296 & 223 & 386 & 12  & 6251 & 1000000 & 1000000 & 0 & 44.48 & 259.88 & 8.72\\
		DMXA0628 & 169 & 280 & 10  & 5236 & 1000000 & 1000000 & 0 & 7.64 & 200.36 & 8.00\\
		DMXA1109 & 343 & 559 & 17  & 73324 & 1000000 & 1000000 & 0 & 27.31 & 307.93 & 7.49\\
		DMXA1304 & 298 & 503 & 10 & 94300 & 1000000 & 1000000 & 0 & 49.84 & 415.11 & 7.40\\
		GAP1307 & 342 & 552 & 17  & 33150 & 1000000 & 1000000 & 0 & 28.42 & 256.28 & 4.19\\
		GAP1500 & 220 & 374 & 17  & 180600 & 1000000 & 1000000 & 0 & 13.78 & 351.57 & 0\\
		MSM0580 & 338 & 541 & 11  & 62086 & 1000000 & 1000000 & 0 & 58.46 & 314.35 & 9.42\\
		MSM1707 & 278 & 478 & 11  & 1000000 & 1000000 & 1000000 & 0.89 & 5.85 & 164.36 & 1.77\\
		MSM1844 & 90 & 135 & 10 & 1960 & 108362 & 1000000 & 0 & 0 & 32.98 & 4.26\\
		MSM4224 & 191 & 302 & 11  & 8964 & 1000000 & 1000000 & 0 & 11.90 & 201.93 & 7.07\\
		MSM4414 & 317 & 476 & 11  & 1000000 & 1000000 & 1000000 & 0.74 & 23.04 & 349.26 & 0\\
		TAQ0891 & 331 & 560 & 10  & 1000000 & 1000000 & 1000000 & 1.57 & 26.02 & 438.87 & 1.57\\
		TAQ0910 & 310 & 514 & 17  & 4255 & 1000000 & 1000000 & 0 & 48.92 & 367.03 & 5.68\\
		TAQ0920 & 122 & 194 & 17  & 3775 & 9786 & 1000000 & 0 & 0 & 92.86 & 2.38\\
		\hline
	\end{tabular}
\end{table*}

As to the fitness evaluation times, MS$^3$-PO has a better performance than both GA and DPSO in 9 out of 18 instances (B1, B2, B4, B7, B10, B11, B13, B14, B16), and in another 6 instances (B5, B6, B8, B9, B15, B17), MS$^3$-PO has a better performance than DPSO. On the other hand, GA has a better performance than DPSO in 15 out of 18 instances (B1, B3, B5-B10, B12-B18). In these instances, the speed of MS$^3$-PO is close to that of GA, and both MS$^3$-PO and GA solve STPG faster than DPSO. As to the errors, MS$^3$-PO has an average error of 0.03$\%$ in these instances, while the average error of GA is 0.38$\%$, the average error of DPSO is 6.71$\%$, and the average error of SPH is 3.35$\%$. Hence, MS$^3$-PO also has an advantage over GA, DPSO and SPH by finding closer approximations to SMTs in these instances. The following conclusion can be made from Table \ref{Comparison of MS$^3$-PO, GA and DPSO in B instances}.

\begin{conclusion}
	MS$^3$-PO has a better performance than GA, DPSO and SPH for solving STPG in the B instances.
\end{conclusion}

To show the effectiveness of our proposed evolutionary computation technique (Equation (\ref{newC1}) \& (\ref{newC2})) to update edge conductivities, each instance of B1-B6 has been simulated for 22 times using different conductivity update equations. In the first 11 times of simulation, the proposed Equation (\ref{newC1}) \& (\ref{newC2}) are used, and we set $\delta=0.2$. In the second 11 times of simulation, the standard Equation (\ref{oldC}) is used. All the other parameters, $\alpha$ and $\mu$, have the same value in different equations. We sequence the first 11 simulation results from the smallest number of fitness evaluation times to the largest number of fitness evaluation times. Then, we sequence the second 11 simulation results in the same way. The median number in each sequence is considered as the average fitness evaluation times required by MS$^3$-PO to find SMTs using the corresponding equation. Two average fitness evaluation times in each instance have been compared in Table \ref{value of new parameter}. In each of B1, B2, B4 and B5, the two average fitness evaluation times have similar values, which means that MS$^3$-PO has close speeds to find SMTs using different conductivity update equations. However, in each of B3 and B6, the average fitness evaluation times using the proposed Equation (\ref{newC1}) \& (\ref{newC2}) is much smaller than the average fitness evaluation times using the standard Equation (\ref{oldC}), which means that MS$^3$-PO finds SMTs much faster using Equation (\ref{newC1}) \& (\ref{newC2}) than using Equation (\ref{oldC}). Therefore, the conclusion below is obtained from this comparison.

\begin{conclusion}
	Our proposed evolutionary computation technique accelerates the optimization process of MS$^3$-PO.
\end{conclusion}

\subsection{Applying MS$^3$-PO to small VLSI design instances}

Here, we apply MS$^3$-PO to small VLSI design instances. Instances with around hundreds of vertices have been selected from SteinLib, and there are 17 instances in total. The comparison results of MS$^3$-PO, GA, DPSO and SPH in these instances are listed in Table \ref{Comparison of MS$^3$-PO, GA and DPSO in VLSI instances}. It can be seen that MS$^3$-PO finds SMTs in 14 out of 17 instances. On the contrary, GA finds SMTs in 2 instances, SPH finds SMTs in 3 instances, and DPSO cannot find an SMT in any instance. Therefore, MS$^3$-PO has an advantage over GA, DPSO and SPH by finding SMTs in more instances.  As to the errors, MS$^3$-PO has an average error of 0.19$\%$ in these instances, while the average error of GA is 28.72$\%$, the average error of DPSO is 279.05$\%$, and the average error of SPH is 4.96$\%$. Thus, MS$^3$-PO also has an advantage over GA, DPSO and SPH by finding closer approximations to SMTs in these instances. The following conclusion can be made from this comparison.

\begin{conclusion}
	MS$^3$-PO has a better performance than GA, DPSO and SPH for solving VLSI design instances with hundreds of vertices.
\end{conclusion}

\begin{table}[!t]
	\vspace{0.1cm}
	\renewcommand{\arraystretch}{1.3}
	\caption{The comparison in large VLSI design instances}
	\label{Comparison of HMS$^3$-PO and SPH in large VLSI instances}
	\centering
	\begin{tabular}{l c c c c c c}
		\hline
		\multicolumn{1}{l|}{\multirow{2}{*}{Instance}} & \multicolumn{1}{c|}{\multirow{2}{*}{$|V|$}} & \multicolumn{1}{c|}{\multirow{2}{*}{$|E|$}} & \multicolumn{1}{c|}{\multirow{2}{*}{$|T|$}}  &\multicolumn{1}{c|}{\multirow{2}{*}{$N$}} & \multicolumn{2}{c}{Error/\%} \\
		\cline{6-7}
		\multicolumn{1}{c|}{}&\multicolumn{1}{c|}{}&\multicolumn{1}{c|}{}&\multicolumn{1}{c|}{}&& \multicolumn{1}{|c}{HMS$^3$-PO} & \multicolumn{1}{|c}{SPH} \\
		\hline
		ALUE2087 & 1244 & 1971 & 34  & 3 & 3.62 & 8.58\\
		ALUE2105 & 1220 & 1858 & 34  & 3 & 3.59 & 3.88\\
		ALUE6951 & 2818 & 4419 & 67 & 10 & 4.32 & 9.22\\
		DIW0234 & 5349 & 10086 & 25  & 5 & 1.65 & 4.71\\
		DIW0445 & 1804 & 3311 & 33  & 5 & 3.89 & 6.82 \\
		DIW0459 & 3636 & 6789 & 25  & 15 & 2.42 & 3.16\\
		DIW0473 & 2213 & 4135 & 25  & 5 & 2.19 & 7.10 \\
		DIW0487 & 2414 & 4386 & 25  & 15 & 4.63 & 7.30\\
		DIW0523 & 1080 & 2015 & 10  & 2 & 0 & 0\\
		GAP2740 & 1196 & 2084 & 14  & 7 & 8.86 & 8.86\\
		GAP3128 & 10393 & 18043 & 104  & 20 & 4.85 & 7.97\\
		MSM0654 & 1290 & 2270 & 10  & 7 & 4.74 & 6.32 \\
		MSM1477 & 1199 & 2078 & 31  & 4 & 4.87 & 10.49 \\
		MSM2525 & 3031 & 5239 & 12  & 5 & 1.01 & 2.40\\
		MSM2601 & 2961 & 5100 & 16  & 10 & 6.53 & 9.03\\
		MSM2705 & 1359 & 2458 & 13  & 3 & 1.82 & 8.68 \\
		MSM2802 & 1709 & 2963 & 18 & 3 & 3.46 & 3.56\\
		MSM3277 & 1704 & 2991 & 12  & 4 & 0 & 1.15 \\
		MSM3727 & 4640 & 8255 & 21  & 8 & 2.18 & 5.09\\
		TAQ0014 & 6466 & 11046 & 128  & 20 & 4.81 & 7.29\\
		TAQ0431 & 1128 & 1905 & 13  & 2 & 5.80 & 8.36\\
		TAQ0751 & 1051 & 1791 & 16  & 3 & 5.86 & 11.18 \\
		\hline
	\end{tabular}
\end{table}

\subsection{Applying HMS$^3$-PO to large VLSI design instances}

Even though MS$^3$-PO has a better performance than GA, DPSO and SPH in VLSI design instances with hundreds of vertices, it is not recommended to use MS$^3$-PO in larger VLSI design instances due to the low speed. In contrast, HMS$^3$-PO can find close approximations to SMT in large graphs in a reasonable time. Considering MS$^3$-PO's good performance in VLSI design instances and the fact that MS$^3$-PO is included in HMS$^3$-PO, HMS$^3$-PO may have a good performance in VLSI design instances as well. To prove it, here, we apply HMS$^3$-PO to large VLSI design instances.

Instances with thousands or tens of thousands of vertices have been selected from SteinLib, and there are 22 instances in total. Given that GA and DPSO need a long time to solve STPG in large graphs, HMS$^3$-PO is only compared with SPH in these instances. The comparison results are listed in Table \ref{Comparison of HMS$^3$-PO and SPH in large VLSI instances}, in which $N$ is the number of partitioned subgraphs in HMS$^3$-PO. In these large VLSI design instances, HMS$^3$-PO has an average error of 3.69$\%$. On the contrary, the average error of SPH is 6.42$\%$. Thus, HMS$^3$-PO provides closer approximations to SMTs than SPH in these instances. Moreover, HMS$^3$-PO will degenerate into SPH when $N$ equals $|T|$. Therefore, HMS$^3$-PO can always find a solution at least as good as the SPH solution in any instance. The following conclusion can be made from this comparison.

\begin{conclusion}
	HMS$^3$-PO can solve STPG in large VLSI design instances with up to tens of thousands of vertices, and it can always provide solutions better than or equal to those of SPH.
\end{conclusion}

\section{Conclusion}
In this paper, we propose two PAs to solve a well-known network optimization problem: the Steiner Tree Problem in Graphs (STPG).  We apply some widely-used artificial and real-world VLSI design instances to evaluate their performances. The experimental results show that: 1) for instances with hundreds of vertices, our first proposed PA can find feasible solutions with an average error of 0.19\%, while the Genetic Algorithm (GA), the Discrete Particle Swarm Optimization (DPSO)  algorithm and a widely-used Steiner tree approximation algorithm: the Shortest Path Heuristic (SPH) algorithm can only find feasible solutions with an average error above 4.96\%; and 2) for larger instances with up to tens of thousands of vertices, where our first proposed PA, GA and DPSO are too slow to be used, our second proposed PA can find feasible solutions with an average error of 3.69\%, while SPH can only find feasible solutions with an average error of 6.42\%. These experimental results indicate that PAs have the potential of computing Steiner trees, and it may be preferable to apply our PAs to solve STPG in some cases.

\balance

\bibliographystyle{ieeetr}
\bibliography{yahuiBibIEEE}

\begin{thebibliography}{10}

\bibitem{imba2000}
T.~Nakagaki, H.~Yamada, and {\'A}.~T{\'o}th, ``Intelligence: Maze-solving by an
  amoeboid organism,'' {\em Nature}, vol.~407, no.~407, pp.~470--470, 2000.

\bibitem{pmer2007}
A.~Adamatzky, ``Physarum machines: encapsulating reaction–diffusion to
  compute spanning tree,'' {\em Naturwissenschaften}, vol.~94, pp.~975--980,
  Jun. 2007.

\bibitem{snsi2004}
T.~Nakagaki, H.~Yamada, and M.~Hara, ``Smart network solutions in an amoeboid
  organism,'' {\em Biophysical chemistry}, vol.~107, pp.~1--5, Jan. 2004.

\bibitem{smeo2012}
A.~Adamatzky and M.~Prokopenko, ``{Slime mould evaluation of Australian
  motorways},'' {\em International Journal of Parallel, Emergent and
  Distributed Systems}, vol.~27, no.~4, pp.~275--295, 2012.

\bibitem{r2a72014}
A.~I. Adamatzky, ``{Route 20, autobahn 7, and slime mold: approximating the
  longest roads in USA and Germany with slime mold on 3-D terrains},'' {\em
  IEEE Transactions on Cybernetics}, vol.~44, no.~1, pp.~126--136, 2014.

\bibitem{fspo2012}
J.~Siriwardana and S.~K. Halgamuge, ``Fast shortest path optimization inspired
  by shuttle streaming of physarum polycephalum,'' in {\em IEEE Congress on
  Evolutionary Computation (CEC)}, pp.~1--8, 2012.

\bibitem{auos2014}
Z.~Zhang, C.~Gao, Y.~Liu, and T.~Qian, ``A universal optimization strategy for
  ant colony optimization algorithms based on the physarum-inspired
  mathematical model,'' {\em Bioinspiration \& biomimetics}, vol.~9, no.~3,
  p.~036006, 2014.

\bibitem{RST2015}
M.~Brazil and M.~Zachariasen, ``{Rectilinear Steiner Trees},'' in {\em Optimal
  Interconnection Trees in the Plane}, pp.~151--218, Springer, 2015.

\bibitem{pspf2002}
E.~Uchoa, M.~Poggi~de Arag{\~a}o, and C.~C. Ribeiro, ``{Preprocessing Steiner
  problems from VLSI layout},'' {\em Networks}, vol.~40, no.~1, pp.~38--50,
  2002.

\bibitem{stst2018}
M.~De~Laere, S.~T. Pham, and P.~De~Causmaecker, ``Solving the steiner tree
  problem in graphs with variable neighborhood descent,'' {\em arXiv preprint
  arXiv:1806.06685}, 2018.

\bibitem{tfha2019}
Y.~Sun, M.~Brazil, D.~Thomas, and S.~Halgamuge, ``The fast heuristic algorithms
  and post-processing techniques to design large and low-cost communication
  networks,'' {\em IEEE/ACM Transactions on Networking}, 2019.

\bibitem{mhnp2019}
Y.~Sun and S.~Halgamuge, ``Minimum-cost heterogeneous node placement in
  wireless sensor networks,'' {\em IEEE Access}, 2019.

\bibitem{apps2016}
Y.~Sun, P.~N. Hameed, K.~Verspoor, and S.~Halgamuge, ``A {Physarum-inspired}
  prize-collecting {Steiner} tree approach to identify subnetworks for drug
  repositioning,'' {\em BMC Systems Biology}, vol.~10, no.~S5, pp.~25--38,
  2016.

\bibitem{tnst2017}
Y.~Sun, C.~Ma, and S.~Halgamuge, ``The node-weighted {Steiner} tree approach to
  identify elements of cancer-related signaling pathways,'' {\em BMC
  Bioinformatics}, vol.~18, no.~S16, pp.~1--13, 2017.

\bibitem{faos2012}
A.~Gubichev and T.~Neumann, ``{Fast approximation of Steiner trees in large
  graphs},'' in {\em Proceedings of the 21st ACM international conference on
  Information and knowledge management}, pp.~1497--1501, Oct. 2012.

\bibitem{sstf2012}
W.~Lee, W.-K. Loh, and M.~M. Sohn, ``{Searching Steiner trees for web graph
  query},'' {\em Computers \& Industrial Engineering}, vol.~62, no.~3,
  pp.~732--739, 2012.

\bibitem{saul2001}
T.~Koch, A.~Martin, and S.~Vo{\ss}, {\em SteinLib: An updated library on
  {Steiner} tree problems in graphs}.
\newblock Steiner Trees in Industry, 2001.

\bibitem{assk2012}
Y.-K. Shih and S.~Parthasarathy, ``A single source k-shortest paths algorithm
  to infer regulatory pathways in a gene network,'' {\em Bioinformatics},
  vol.~28, no.~12, pp.~i49--i58, 2012.

\bibitem{pdim2010}
K.~Faust, P.~Dupont, J.~Callut, and J.~Van~Helden, ``Pathway discovery in
  metabolic networks by subgraph extraction,'' {\em Bioinformatics}, vol.~26,
  no.~9, pp.~1211--1218, 2010.

\bibitem{omsf2004}
T.~Nakagaki, R.~Kobayashi, Y.~Nishiura, and T.~Ueda, ``{Obtaining multiple
  separate food sources: behavioural intelligence in the Physarum
  plasmodium},'' {\em Proceedings of the Royal Society of London B: Biological
  Sciences}, vol.~271, pp.~2305--2310, Nov. 2004.

\bibitem{faip2008}
A.~Tero, K.~Yumiki, R.~Kobayashi, T.~Saigusa, and T.~Nakagaki, ``Flow-network
  adaptation in physarum amoebae,'' {\em Theory in Biosciences}, vol.~127,
  pp.~89--94, Apr. 2008.

\bibitem{amib2010}
A.~Tero, T.~Nakagaki, K.~Toyabe, K.~Yumiki, and R.~Kobayashi, ``A method
  inspired by {Physarum} for solving the {Steiner} problem,'' {\em
  International Journal of Unconventional Computing}, vol.~6, pp.~109--123,
  2010.

\bibitem{poab2015}
L.~Liu, Y.~Song, H.~Zhang, H.~Ma, and A.~V. Vasilakos, ``Physarum optimization:
  A biology-inspired algorithm for the {Steiner} tree problem in networks,''
  {\em IEEE Transactions on Computers}, vol.~64, pp.~819--832, Mar. 2015.

\bibitem{faib2016}
Y.~Sun and S.~Halgamuge, ``{Fast algorithms inspired by Physarum polycephalum
  for node weighted Steiner tree problem with multiple terminals},'' in {\em
  IEEE Congress on Evolutionary Computation}, pp.~3254--3260, IEEE, 2016.

\bibitem{aasf1980}
H.~Takahashi and A.~Matsuyama, ``An approximate solution for the {Steiner}
  problem in graphs,'' {\em Math. Japonica}, vol.~24, no.~6, pp.~573--577,
  1980.

\bibitem{pccs2012}
V.~Bonifaci, K.~Mehlhorn, and G.~Varma, ``Physarum can compute shortest
  paths,'' {\em Journal of theoretical biology}, vol.~309, pp.~121--133, 2012.

\bibitem{psab2006}
A.~Tero, R.~Kobayashi, and T.~Nakagaki, ``Physarum solver: a biologically
  inspired method of road-network navigation,'' {\em Physica A: Statistical
  Mechanics and its Applications}, vol.~363, pp.~115--119, Apr. 2006.

\bibitem{ammf2007}
A.~Tero, R.~Kobayashi, and T.~Nakagaki, ``A mathematical model for adaptive
  transport network in path finding by true slime mold,'' {\em Journal of
  theoretical biology}, vol.~244, no.~4, pp.~553--564, 2007.

\bibitem{sbot2001}
T.~Nakagaki, ``Smart behavior of true slime mold in a labyrinth,'' {\em
  Research in Microbiology}, vol.~152, no.~9, pp.~767--770, 2001.

\bibitem{plan2013}
T.~Sch{\"o}n, {\em Physarum Learner: a novel structure learning algorithm for
  Bayesian Networks inspired by Physarum polycephalum}.
\newblock PhD thesis, 2013.

\bibitem{toir2011}
S.~Watanabe, A.~Tero, A.~Takamatsu, and T.~Nakagaki, ``Traffic optimization in
  railroad networks using an algorithm mimicking an amoeba-like organism,
  physarum plasmodium,'' {\em Biosystems}, vol.~105, no.~3, pp.~225--232, 2011.

\bibitem{caoc2008}
T.~Nakagaki, A.~Tero, R.~Kobayashi, I.~Onishi, and T.~Miyaji, ``Computational
  ability of cells based on cell dynamics and adaptability,'' {\em New
  Generation Computing}, vol.~27, pp.~57--81, Nov. 2008.

\bibitem{otso2015}
M.~Caleffi, I.~F. Akyildiz, and L.~Paura, ``{{On the solution of the Steiner
  tree NP-hard problem via physarum BioNetwork}},'' {\em IEEE/ACM Transactions
  on Networking}, vol.~23, no.~4, pp.~1092--1106, 2015.

\bibitem{aphf2014}
M.~Leitner, I.~Ljubi{\'c}, M.~Luipersbeck, and M.~Resch, ``A partition-based
  heuristic for the {Steiner} tree problem in large graphs,'' {\em Hybrid
  Metaheuristics}, pp.~56--70, 2014.

\bibitem{stgs1993}
A.~Kapsalis, V.~J. Rayward-Smith, and G.~D. Smith, ``Solving the graphical
  {Steiner} tree problem using genetic algorithms,'' {\em Journal of the
  Operational Research Society}, vol.~44, pp.~397--406, Apr. 1993.

\bibitem{apso2010}
X.~Ma and Q.~Liu, ``{A particle swarm optimization for Steiner tree problem},''
  in {\em IEEE International Conference on Natural Computation (ICNC)}, vol.~5,
  pp.~2561--2565, 2010.

\bibitem{stmf2013}
A.~Sadeghi and H.~Fr{\"o}hlich, ``Steiner tree methods for optimal sub-network
  identification: an empirical study,'' {\em BMC bioinformatics}, vol.~14,
  no.~1, p.~144, 2013.

\bibitem{phft1992}
P.~Winter and J.~M. Smith, ``{Path-distance heuristics for the Steiner problem
  in undirected networks},'' {\em Algorithmica}, vol.~7, no.~1-6, pp.~309--327,
  1992.

\bibitem{ofsv1986}
V.~J. Rayward-Smith and A.~Clare, ``On finding {Steiner} vertices,'' {\em
  Networks}, vol.~16, pp.~283--294, Aug. 1986.

\bibitem{asaf1989}
J.~E. Beasley, ``{An SST-based algorithm for the steiner problem in graphs},''
  {\em Networks}, vol.~19, pp.~1--16, Jan. 1989.

\bibitem{boaa2014}
T.~Weise, R.~Chiong, J.~Lassig, K.~Tang, S.~Tsutsui, W.~Chen, Z.~Michalewicz,
  and X.~Yao, ``Benchmarking optimization algorithms: An open source framework
  for the traveling salesman problem,'' {\em IEEE Computational Intelligence
  Magazine}, vol.~9, no.~3, pp.~40--52, 2014.

\end{thebibliography}

\end{document}